\documentstyle[12pt,epsfig,overcite]{article}
\def\nostrocostrutto#1\over#2{\mathrel{\mathop{\kern 0pt \rlap 
  {\raise.2ex\hbox{$#1$}}}
  \lower.9ex\hbox{\kern-.190em $#2$}}}

\newcommand{\be}{\begin{equation}}
\newcommand{\ee}{\end{equation}}
\newcommand{\ba}{\begin{eqnarray}}
\newcommand{\ea}{\end{eqnarray}}

\begin{document}

\null
\rightline{MPI-PhT/96-36} 
\rightline{May 1996} 
\vspace{2cm}

 \centerline{\Large\bf Perturbative QCD approach and} 
\vspace{0.1cm} 
\centerline{\Large\bf particle energy spectra%
\footnote{to be
published in the Proceedings of the XXXI$^{st}$ Rencontre de Moriond, ``QCD
and High Energy Interactions'', Les Arcs, March 23$^{rd}$-30$^{th}$, 1996, 
Ed. J. Tran Th\^anh V\^an, Editions Fronti\`eres, Paris-CEDEX} 
 } 
\vspace{1.0cm} 
\centerline{SERGIO LUPIA}
\vspace{1.0cm}
\centerline{\it Max-Planck-Institut f\"ur 
Physik} 
\centerline{\it (Werner-Heisenberg-Institut)} 
\centerline{\it F\"ohringer Ring 6, D-80805 M\"unchen, Germany}
\centerline{E-mail: lupia@mppmu.mpg.de}

\vspace{2.0cm} 
\begin{abstract}
\noindent 
The Modified Leading Log Approximation of QCD and the local parton
hadron duality picture are shown to 
give a fully satisfactory description of data on 
charged particle energy spectra in $e^+e^-$ annihilation. The analysis 
of data in terms of moments points out the running of $\alpha_s$ 
both at very low $cms$ energy (${\cal O}$(3 GeV))  and at very low 
particle energy (${\cal O}$(200-500 MeV)). 
The study of the universal soft limit of energy spectra 
at low particle energy 
is proposed as a test of QCD coherence.
The extension of this approach at
identified particles' spectra is discussed.
\end{abstract}

\newpage 

\baselineskip=18pt


The Modified Leading Log Approximation (MLLA) 
of QCD\cite{DKMTbook} plus the Local Parton Hadron Duality (LPHD)\cite{LPHD} 
as hadronization prescription 
give a quite successful description of production process of 
soft and semihard particles (for a recent  discussion see e.g.~\cite{KO}).  
Predictions on the single particle inclusive momentum spectrum, the 
well-known ``hump-backed plateau'',  is  one of the best successes 
of this approach. 
However, the study of the spectrum alone does not allow to see to what extent 
experimental data reveals peculiar features of the theory, like for 
instance coherence and the running of the coupling. 
In this short note, new alternative analyses  more
sensitive than the spectrum itself to the different aspects of the theory 
are discussed. Due to the limited amount of space, I just give the flavour of
our studies and I refer to the original 
literature\cite{lo,klo,loprep} for more detailed presentations. 


MLLA predictions\cite{dkt5}   for the moments 
of single particle inclusive energy spectrum 
have been compared in \cite{lo} 
with moments extracted from experimental spectra. 
Good agreement between experimental data and theoretical predictions of MLLA
with running coupling has been found in a wide $cms$ energy range 
from 3 up to LEP energy. 
The model with fixed coupling can reproduce the multiplicity
and the slope of the first moment only. 
The two models are particularly different at low $cms$ energies, where the
effect of the running coupling is large. 
Notice that the model with fixed coupling cannot describe the data 
even if one relaxes the normalization at thresholds\cite{loprep}. 
The sensitivity of moments to the running of the 
QCD coupling has then been established; 
the validity of the description at low $cms$ energy is particularly 
interesting. In this respect, it is important to point out that 
further tests of this picture
could be performed at HERA (in the Breit frame), where many data points at
different $Q^2$  could be obtained within the same experiments. 
Going to larger $cms$ energies, the extrapolation of theoretical predictions
have been shown to be successful at  LEP-1.5 $cms$
energies\cite{klo}; the same analysis could be repeated at the forthcoming
LEP-2.
 

\begin{figure}
\vfill \begin{minipage}{.48\linewidth}
          \begin{center}
   \mbox{
\mbox{\epsfig{file=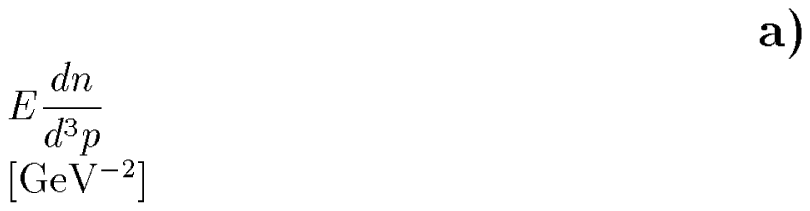,bbllx=5.2cm,%
bblly=17.5cm,bburx=5.4cm,bbury=25.cm}}
\mbox{\epsfig{file=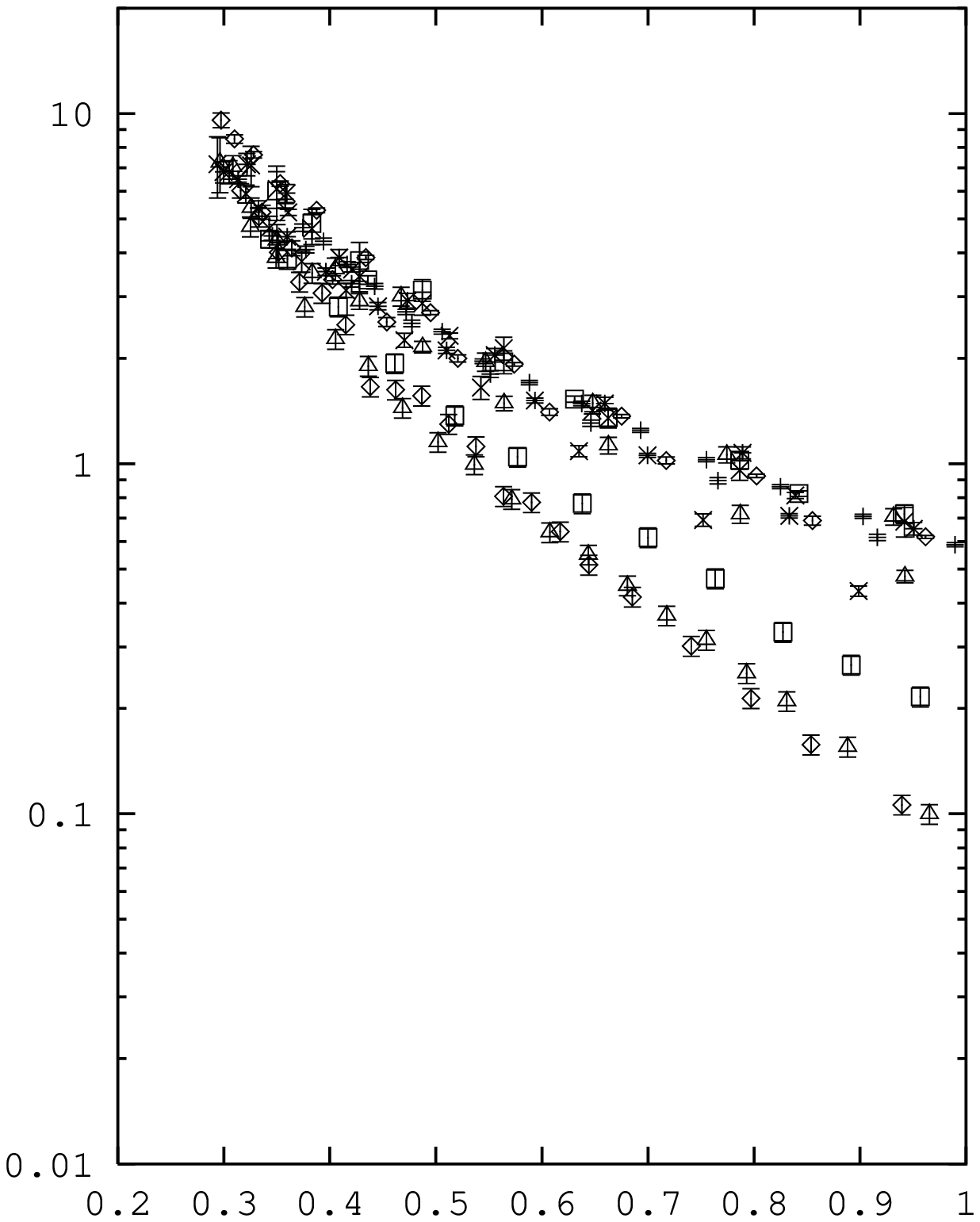,%
width=.72\linewidth,bbllx=5.5cm,bblly=2.5cm,bburx=15.cm,bbury=18.cm}}
  }        \end{center}
\vspace{-0.3cm}
\centerline{$\quad \qquad E$ [GeV]}
      \end{minipage}\hfill
      \begin{minipage}{.48\linewidth}
          \begin{center}
   \mbox{
\mbox{\epsfig{file=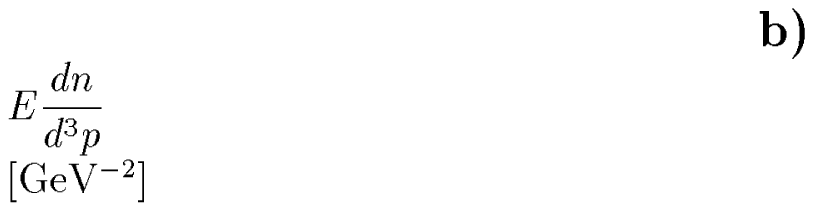,bbllx=5.2cm,%
bblly=17.5cm,bburx=5.4cm,bbury=25.cm}}
\mbox{\epsfig{file=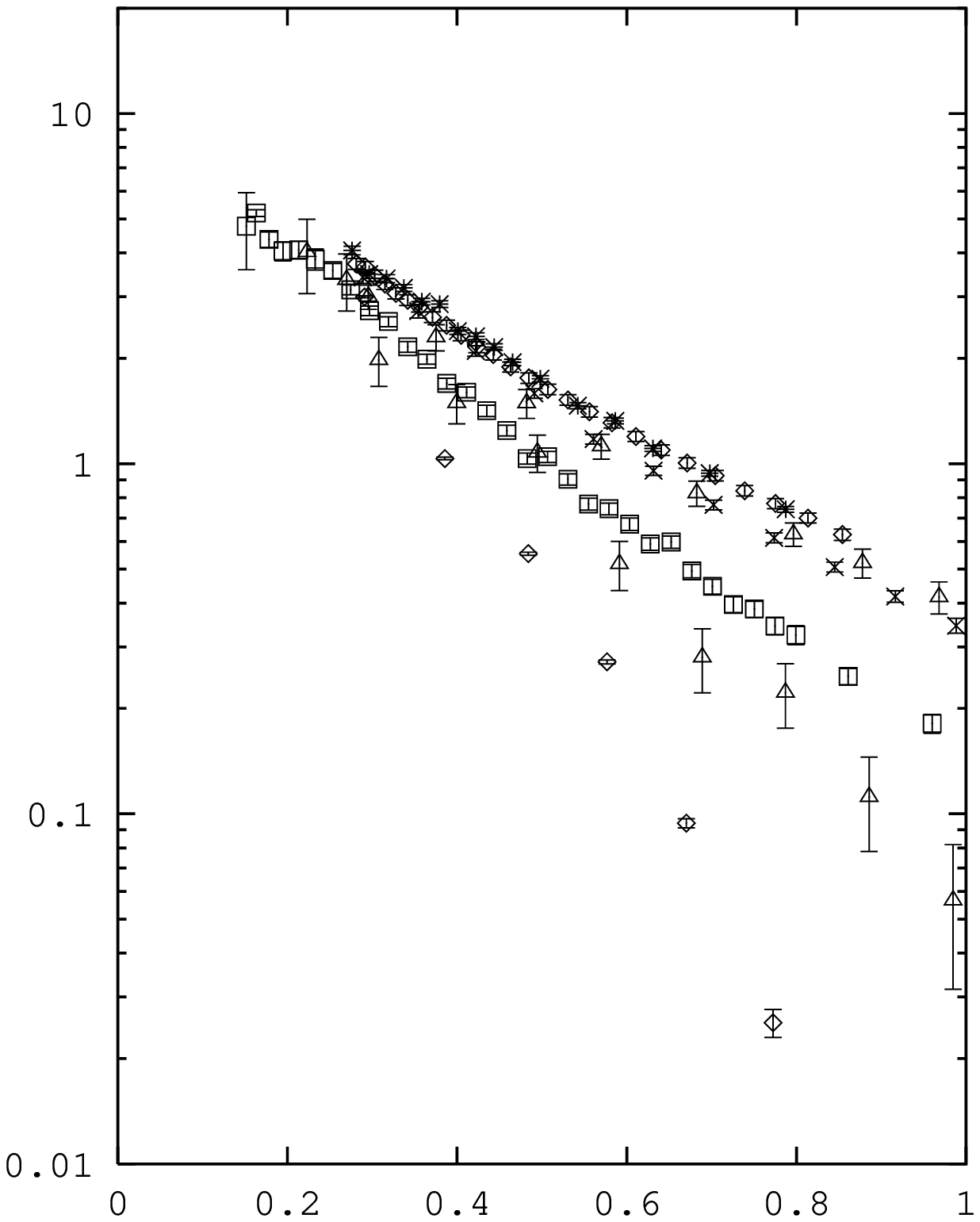,%
width=.72\linewidth,bbllx=5.5cm,bblly=2.5cm,bburx=15.cm,bbury=18.cm}}
  }        \end{center}
\vspace{-0.3cm}
\centerline{$\quad \qquad E$ [GeV]}
      \end{minipage}
\caption{{\bf a)}: Invariant distribution $E dn/d^3p$ for charged particle 
as a function of the particle energy $E$ with $Q_0$ = 270 MeV. Data extracted 
from $^{10,11)}$. 
{\bf b)}: same as in {\bf a)} but for charged pions with
$Q_0$ = $m_{\pi}$; data from $^{12,11)}$.} 
\end{figure} 

Let us now look at the invariant cross section $E dn/d^3p$ 
at particle energy $E$ of few hundreds MeV. 
QCD coherence  predicts that very soft particles do not
multiply\cite{DKMTbook};  
since they have a long wavelength, they should see  the total 
initial charge only.   Particle production  should then 
 be independent of $cms$ energy  for low energy particles~\cite{vakcar}.  
 In 
 addition,  $E dn/d^3p$ should go to 
 a finite limit for $E \to m_h$; 
 both the scaling with $cms$ energy and the existence of a soft limit
 can be derived analytically in Double Log
Approximation\cite{lo}. 
As shown in Figure~1a, both  effects  are seen in experimental data for charged
particle spectra in a wide range of $cms$ energies; 
the approximate scaling with c.m. energy 
coincides with the scaling observed by the OPAL
Collaboration\cite{refcharged} in the variable $\ln p$ at  small momenta. 
Let us also notice that MLLA predictions (here not shown, 
see~\cite{lo,loprep}) 
not only show a similar behavior, but can reproduce also 
quantitatively the experimental data down to very low particle energies. 
On the other hand, 
by switching off the running of the QCD coupling, MLLA predictions 
show  a flatter slope far away from experimental behavior. 
Thus, the running of the QCD coupling is visible in the particle spectra
even in the energy range of a  few hundred MeV. In addition, the approximate
scaling law and the existence of a finite soft limit strongly support QCD
coherence. 
As shown in Figure~1b, the approximate scaling and the finite limit are seen
also in inclusive 
pion spectra at different $cms$ energies; more detailed studies 
on  $\pi, K$ and $p$ 
inclusive spectra are in progress\cite{loprep}.


In conclusion, 
good agreement of experimental data on moments of inclusive energy spectra 
in $e^+e^-$ collisions with predictions of 
MLLA with running $\alpha_s$ plus LPHD (2 free parameters) 
for $\sqrt{s}$ ranging from about 3 GeV to LEP-1.5 energy has been found.
MLLA with fixed $\alpha_s$ is  unable to reproduce moments both with and
without absolute normalization at threshold. 
Data on the invariant cross section $E\ dn/d^3p$ 
give evidence for the existence of a finite soft limit 
and approximate scaling  at different $cms$ energies 
at particle energy of few hundreds MeV,  as suggested by QCD coherence. 
MLLA with  running $\alpha_s$ reproduces this effect.
Direct evidence for running $\alpha_s$ effects in hadronic spectra, 
test of coherence effects   and further evidence for LPHD  
 down to low $cms$ energies and to low particle energies have been achieved.

\vspace{0.6cm}

{\it\noindent 
I thank Wolfgang Ochs and Valery A. Khoze for discussion 
and collaboration on the subjects of this
talk. I thank J. Tran Th\^anh V\^an, A. Capella 
 and the Organizing Committee for the 
nice and stimulating atmosphere created at the Conference. 
}

\end{document}